 \documentstyle[12pt]{article}

 \let\m=\mu

\let\la=\label
  
 \def\bd{\begin{document}} \def\ed{\end{document}}
\def\ds{\documentstyle} \let\fr=\frac \let\bl=\bigl \let\br=\bigr
\let\Br=\Bigr \let\Bl=\Bigl
\let\bm=\bibitem
\let\na=\nabla
\let\pa=\partial \let\ov=\overline
\newcommand{\be}{\begin{equation}}
\newcommand{\ee}{\end{equation}}
\def\ba{\begin{array}}
\def\ea{\end{array}}
\newcommand{\ho}[1]{$\, ^{#1}$}
\newcommand{\hoch}[1]{$\, ^{#1}$}
\newcommand{\bea}{\begin{eqnarray}}
\newcommand{\eea}{\end{eqnarray}}
\newcommand{\ra}{\rightarrow}
\newcommand{\lra}{\longrightarrow}
\newcommand{\Lra}{\Leftrightarrow}
\newcommand{\ap}{\alpha^\prime}
\newcommand{\bp}{\beta^\prime}
\newcommand{\tr}{{\rm tr} }
\newcommand{\Tr}{{\rm Tr} }
\newcommand{\NPge}{Nucl. Phys. }
\newcommand{\tamphys}{\it
Theoretical Physics, Blackett Laboratory, Imperial College London,\\
 London SW7 2AZ, United Kingdom

\&\\
Mathematical Institute, Andrew Wiles Building, University of Oxford,
 \\
Woodstock Road, Radcliffe Observatory Quarter, \\
Oxford OX2 6GG, United Kingdom\\

\&\\
Institute for Quantum Science and Engineering and Hagler Institute for Advanced Study, Texas A\&M University, College Station, TX, 77840, USA}

\begin{document}
\hfill{ Imperial-TP-2020-MJD-01}

\vspace{12pt}

\begin{center}
{ \large {\bf Weyl, Pontryagin, Euler, Eguchi and Freund}}

\vspace{24pt}

M. J. Duff

\vspace{5pt}


{\tamphys}

\vspace{15pt}

IN MEMORY OF PETER FREUND
\vspace{30pt}

ABSTRACT
\end{center}
In a September 1976 PRL Eguchi and Freund considered two topological invariants: the Pontryagin number $P \sim \int d^4x \sqrt{g}R^* R$  and the Euler number $\chi \sim \int d^4x \sqrt{g}R^* R^*$ and posed the question: to what anomalies do they contribute? They found that $P$ appears in the integrated divergence of the axial fermion number current, thus providing a novel topological interpretation of the anomaly found by Kimura in 1969 and Delbourgo and Salam in 1972.  However, they found no analogous role  for $\chi$. This provoked my interest and, drawing on my April 1976 paper with Deser and Isham on gravitational Weyl anomalies, I was able to show that for Conformal Field Theories   the trace of the stress tensor depends on just two constants:
\[
g^{\mu\nu}\langle T_{\mu\nu}\rangle=\frac{1}{(4\pi)^2}(cF-aG)
\]
where $F$ is the square of the Weyl tensor and $\int d^4x\sqrt{g} G/(4\pi)^2$ is the Euler number.                                                                                                                                                                       For free CFTs with $N_s$  massless fields of spin $s$
\[
720c=6N_0 + 18N_{1/2} + 72 N_1~~~~ 720a=2N_0 + 11N_{1/2} + 124N_1  
\]

\pagebreak

\setcounter{page}{1}

\section{Freund}
Peter Freund ranks highly on the list of physicists who have influenced my work, especially in the realm of Kaluza-Klein supergravity \cite{Duff:1986hr}, but here I have chosen to recollect an older source of inspiration, namely his paper with Eguchi\footnote{Sadly, Tohru Eguchi also died recently.} on topological invariants and anomalies \cite{Eguchi:1976db}.

\section{Weyl}

Following the discovery of the gravitational Weyl anomaly by Capper and myself \cite{Capper:1974ic} in 1974,  Deser, Isham and I decided in April 1976  to write down the most general form
of the trace of the
energy-momentum tensor in various dimensions \cite{Deser:1976yx}.  By general
covariance and dimensional
analysis, it must take the following form: 
\begin{itemize}
\item{$D=2$},
\begin{equation}
g^{\alpha\beta}\langle T_{\alpha\beta} \rangle = {\sf c}R,
\label{n=2}
\end{equation}
\item{$D=4$},
\begin{equation}
g^{\alpha\beta}\langle T_{\alpha\beta} \rangle = \alpha R^2+\beta R_{\mu\nu}R^{\mu\nu}+
\gamma
R_{\mu\nu\rho\sigma}R^{\mu\nu\rho\sigma}+\delta{}\raisebox{.7ex}{\fbox{}} R
\label{n=4}
\end{equation}
\item{$D\geq 6$}
\begin{equation}
g^{\alpha\beta}\langle T_{\alpha\beta} \rangle\sim(\textrm{Riem})^{D/2}+...
\end{equation}
\end{itemize}
where ${\sf c}, \alpha, \beta, \gamma, \delta$ are constants.  (At one-loop, and
ignoring boundary terms, there is no anomaly for $D$ odd).

The significance of my paper with Deser and Isham was
to demonstrate that, in addition to scheme-dependent terms such as ${}\raisebox{.7ex}{\fbox{}} R$ which can be removed  by the addition of finite local counterterms such as $R^2$, there are scheme-independent terms such as $\alpha R^2+\beta R_{\mu\nu}R^{\mu\nu}+
\gamma
R_{\mu\nu\rho\sigma}R^{\mu\nu\rho\sigma}$ which cannot, thus laying  to rest any lingering doubts about the
inevitability of
Weyl anomalies. As is well-known, the 1PI effective action arising from closed loops of massless particles is non-local. The title of the paper {\it Non-local
Conformal Anomalies} was chosen to emphasize that although the trace of the stress tensor and infinite counterterms are local, (for example $C_{\mu\nu\rho\sigma}C^{\mu\nu\rho\sigma}$),  the part of the finite effective action responsible for the (curvature)$^2$ anomalies is  not\footnote{This means, in particular,
that Starobinsky's original model of cosmic inflation \cite{Starobinsky:1980te}, which was driven by the Weyl anomaly, involved a non-local lagrangian and not the local $R+R^2$ as is often stated. See also \cite{Hawking:2000bb,Pelinson:2002ef}.} (for example $C_{\mu\nu\rho\sigma}~ \ln ~{}\raisebox{.7ex}{\fbox{}}~ C^{\mu\nu\rho\sigma}$).  One may construct a local action involving extra scalar fields which yields the non-local action after  integrating out the scalars \cite{Mazur:2001aa}. However, the exact form of the non-localities is still a matter of debate. See \cite{Fradkin:1983tg,Riegert:1984kt,Barvinsky:1994cg,Deser:1996na,Deser:1999zv,Mazur:2001aa,Meissner:2007xv,Meissner:2018hmx,
Kuzenko:2019vvi}

\section {Eguchi and Freund}
The scalar terms of order $D/2$ in the curvature which appear in the
$D$-dimensional
gravitational trace anomaly are reminiscent of the pseudoscalar terms of order
$D/2$ in the
curvature which appear in the $D$-dimensional gravitational axial anomaly as
calculated for massive Dirac spinors $\psi$ in $D=4$ by Kimura \cite{Kimura:1969} and by Delbourgo
and Salam \cite{Delbourgo:1972xb} 

 \be
 \partial_{\mu} (\sqrt{g}J^ {\mu}{}_ 5)=2m\sqrt{g}J_5 -\frac{1}{384\pi^2}{}\epsilon^{\mu\nu\rho\sigma}R_{\alpha\beta\mu\nu}R^{\alpha\beta}{}_{\rho\sigma} 
 \label{kds}
\ee
 where
\be
J^{\mu}{}_5={\bar \psi}\gamma^{\mu}\gamma_5\psi
\ee
 and
 \be
J_5={\bar \psi}\gamma_5\psi
\ee

 I was musing on this in September 1976 when I saw a paper by Eguchi and Freund  in PRL \cite{Eguchi:1976db} on the then new
and exciting topic of
gravitational instantons.  They considered two topological invariants; the Pontryagin number 
\be
P=\frac{1}{(4\pi)^2} \int d^4 x \sqrt{g}R^*R 
\ee
 and the Euler number 
  \be
  \chi =\frac{1}{2}\frac{1}{(4\pi)^2}\int d^4 x \sqrt{g}R^*R^*
  \ee
  where
  \be
  *R^{\mu\nu}{}_{\alpha\beta} =\frac{1}{2}\epsilon^{\mu\nu\rho\sigma}R_{\alpha\beta\mu\nu}
  \ee
 and posed the question: to what anomalies do they contribute? They found that $P$ appears in the divergence of the axial fermion number current 
   \be
\int d^4x \partial _{\mu}(\sqrt{g} J^ {\mu}{}_5)=\frac{N}{12}P
   \ee
in the case of $N$ massless Dirac fermions thus providing a topological interpretation of the results of Kimura and Delbourgo and Salam.  For $\chi$ however, they say ``We now consider the other topological invariant of the gravitational field, the Euler-Poincare characteristic $\chi$. Its density $R^* R^*$ does not seem to lead to any anomalies.''

I therefore wrote a short note to PRL \cite{Duff:1976nj} relating $\chi$ to the
integrated trace
anomaly.  As described in  \cite{Polyakov:1981rd}, this result was later to prove
important in the
two-dimensional context of string theory, 
\begin{equation}
\frac{1}{ 4{\pi}}\int d^2x\sqrt{g}g^{\alpha\beta}\langle T_{\alpha\beta}\rangle = {\sf  c}{\chi}.
\label{n=2euler}
\end{equation}
where the worldsheet Euler number is related to the genus  $g$ of the Riemann surface 
\begin{equation}
\chi=\frac{1}{4\pi}\int d^2x\sqrt{g}R=2-2g
\label{chi}
\end{equation}
Unfortunately, the referee's vision did not extend
that far and the paper was rejected.  Rather than  resubmit it, I decided to
incorporate the results
into a larger paper \cite{Duff:1977ay} which re-examined the Weyl anomaly in the
light of its applications
to the Hawking effect, to gravitational instantons, to asymptotic freedom and
Weinberg's asymptotic
safety.
\section{The $c$ and $a$ coefficients}

In the process, I discovered that  for Conformal Field Theories (CFTs) in $D=4$ the constants $\alpha$, $\beta$, $\gamma$ and $\delta$
are not all independent but obey the constraints
\begin{equation}
4\alpha+\beta=\alpha-\gamma=-\delta
\label{constraint}
\end{equation}
so that  the trace of the stress tensor depends  on just two constants\footnote{In the notation of \cite{Duff:1977ay} $(4\pi)^{2}b=c$, $(4\pi)^{2}b'=-a$ and $N_{1/2}$ counts the number of 4-component spinors. The letters $c$ and $a$ and the nomenclature of central charges may be found in \cite{Anselmi:1997am} along with a discussion of the anomaly supermultiplet.}:
\be
g^{\mu\nu} \langle T_{\mu\nu} \rangle=\frac{1}{(4\pi)^2}(cF-aG)
\la{canda}
\ee
where
\begin{equation}
F=C^{\mu\nu\rho\sigma}C_{\mu\nu\rho\sigma},
\label{weyltensor}
\end{equation}
 $C_{\mu\nu\rho\sigma}$ is the Weyl tensor, and $G$ is proportional to the
Euler number density
\begin{equation}
G=R^*{}_{\mu\nu\rho\sigma}R^*{}^{\mu\nu\rho\sigma}.
\label{eulerdensity}
\end{equation}
(We have ignored the scheme-dependent  $\raisebox{.7ex}{\fbox{}} R$ term, but see \cite{Prochazka:2017pfa})
Moreover, for free  CFTs with $N_s$ massless fields of spin $s$,
\be
720c=6N_0 + 18N_{1/2} + 72N_1,~~~~ 720a=2N_0 + 11N_{1/2} + 124N_1  .
\ee
Here $N_{1/2}$ counts the number of 2-component spinors; 4-component spinors contribute twice as much. Note the
inequalities on the ratio $a/c$ 
\be
\frac{31}{18 }\geq \frac{a}{c}\geq\frac{1}{3}
\ee
where the upper and lower bounds are saturated by a single vector and a single scalar, respectively.  Remarkably, these bounds continue to hold true when the CFT is interacting \cite{Hofman:2008ar}.

The Weyl anomaly acquires a new significance when placed in the context of supersymmetry. In particular, Ferrara and Zumino \cite{Ferrara:1974pz} showed that the trace of the stress tensor $T_\mu{}^\mu$, the divergence of the axial current $\partial_\mu J^{\mμ 5}$, and the gamma trace of the spinor current $\gamma_\mu S^{\mu}$ form a scalar supermultiplet. See  \cite{Anselmi:1997am} . Table 1 shows the values of $a$ and $c$ for various supermultiplets.

 The  Weyl anomaly in four-derivative theories, such as  {\it Conformal Gravity} and {\it Conformal Supergravity} \cite{Fradkin:1983tg}, also takes the form (\ref{canda}) with suitably corrected $c$ and $a$. So does
the non-perturbative {\it holographic} Weyl anomaly \cite{Henningson:1998gx,Graham:1999pm}, which plays a vital part in the  AdS/CFT correspondence.  In the large $N$ limit $c=a$. See also the earlier work in \cite{Liu:1998bu}.

See for example \cite{Henningson:1998gx,Bastianelli:2000hi,Astaneh:2015tea} for the $D=6$ anomaly .
 
  \begin{center}
  \begin{table}
   \begin{tabular}{|l||c|c|cc|c|c|} \hline      
    Fields                         &$ a      $&$ c  $ &$a/c$&\\ 
    \hline
     ${\cal N}=0$ spin $0$                 &$ 1/360  $&$ 1/120$&$1/3$&\\
    ${\cal N}=0$  spin $1/2$              &$ 11/720 $&$ 1/40$ &$11/18$&\\
     ${\cal N}=0$ spin $1$                 &$ 31/180 $&$ 1/10 $&$31/18$&\\    
  \hline
    ${\cal N}=1$ chiral multiplet  &$ 1/48   $&$ 1/24$&$1/2$&\\ 
    ${\cal N}=1$ vector multiplet  &$ 3/16   $&$ 1/8$&$3/2$&\\
       \hline
    ${\cal N}=2$ hyper multiplet   &$ 1/24   $&$ 1/12 $&$1/2$& \\ 
    ${\cal N}=2$ vector multiplet  &$ 5/24   $&$ 1/6$&$5/4$&\\
       \hline
    ${\cal N}=4$ vector multiplet  &$ 1/4    $&$ 1/4 $&$1$& \\
    \hline
   \end{tabular}
\label{aandc}
\caption{The central charges $a$ and $c$ for CFTs}
\end{table}
\end{center}

\section{Comments and caveats}

\begin{itemize}
\item The integrated anomaly reduces to the pure Euler form in the important special cases of Ricci flat ($F=G$)
\be
g^{\mu\nu} \langle T_{\mu\nu} \rangle=\frac{1}{(4\pi)^2}(c-a)G
\la{rflat}
\ee
 and conformally flat ($F=0$) 
\be
g^{\mu\nu} \langle T_{\mu\nu} \rangle=-\frac{1}{(4\pi)^2}aG
\la{cflat}
\ee
spacetimes.

\item
Note the absence of an $R^2$ term in (\ref{canda}).
This result was later
rederived using the Wess-Zumino consistency conditions
\cite{Bonora:1983ff}.
\item
The constants
${\sf c}$, $a$, $c$ 
are also  those which determine the counterterms

\begin{equation}
\Delta L=\frac{1}{\epsilon}{\sf c}\sqrt{g}R
\label{n=2counterterm}
\end{equation}
\begin{equation}
\Delta L=\frac{1}{\epsilon}\frac{1}{(4\pi)^2}\sqrt{g}(cF-aG)
\label{n=4counterterm}
\end{equation}
at the one-loop level. 
The Euler number counterterms were previously ignored on the grounds that they are total
divergences \cite{Hooft:1974bx,Capper:1974ic}, but will
nevertheless contribute on the spacetimes of non-trivial topology demanded by gravitational instantons. 
\item
By calculating the two-point function 
\bea
\Pi_{\mu\nu\rho\sigma}(p)&=&\int d^{D}xe^{ipx} \langle T_{\mu\nu}(x)T_{\rho\sigma}(0)\rangle |_{_{g_{\mu\nu}=\eta_{\mu\nu}}}\\
&=&\frac{1}{\epsilon}\frac{c}{(4\pi)^2}S^{(2)}{}_{\mu\nu\rho\sigma}+ finite
\label{pi}
\eea
using dimensional regularization ($\epsilon=D-4$) where
\be 
S^{(2)}{}_{{\mu\nu\rho\sigma}}=\frac{1}{2}(X_{\mu\rho} X_{\nu\sigma}+X_{\nu\rho} X_{\mu\sigma})-\frac{1}{3}X_{\mu\nu} X_{\rho\sigma}
\ee
\be
X_{\mu\nu}=\eta_{\mu\nu}p^2-p_{\mu}p_{\nu}\\
\ee
the earlier papers by Capper and myself \cite{Capper:1973mv,Capper:1973bk,Capper:1974ic,Capper:1974ed,Capper:1975ig} determined $c$, the coefficient of the $(\textrm{Weyl})^2$ counterterm but not $a$, the Gauss-Bonnet term which requires the three-point function. See \cite{Duff:1977ay,Duff:1993wm} for a list of references for $a$ calculations.
The two-point function is sufficient to calculate CFT loop corrections to the Schwarzschild solution and hence Newton's law  \cite{Duff:1974ud,Duff:2000mt}, 
\be
V(r)= \frac{G_{}M}{r}\biggl
(1+\frac{8cG_{}}{3\pi r^{2}} \biggr).
\label{AdS}
\ee
For ${\cal N}=4$ Yang-Mills $c=1/4$ and we recover the 
Randall-Sundrum \cite{Randall:1999vf} braneworld result.
Similarly the graviton mass in the Karch-Randall \cite{Karch:2000ct} braneworld is
\be
M^2=\frac{6cG}{\pi L^4},
\ee
where $L$ is the $AdS_4$ radius.
\item

The $- {\it euler}+ {\it conformal} $ structure of the anomaly (\ref{canda}) may be generalized to arbitrary even dimensions and are labelled Type A and Type B respectively \cite{Deser:1993yx}. 
\be
g^{\mu\nu} \langle T_{\mu\nu} \rangle=-(-1)^{d/2}aE_d+\sum_ic^iI_i
\la{csanda}
\ee
where $E_d$ is the Euler density arising from scale-free contributions to the effective gravitational action and the $I_i$ are local conformal scalar polynomials involving powers of the Weyl tensor and its derivatives; their number increases rapidly with dimension. See \cite{Cappelli:2000fe} for the $a$ coefficients in arbitrary dimensions.
\item 
The two dimensional anomaly is purely Type A, leading to the conjecture that the  {\sf c}-theorem on renormalisation group flow  in $D=2$ \cite{Zamolodchikov:1986gt} would generalize to an $a$-theorem in $D=4$ \cite{Cardy:1988cwa,Osborn:1989td,Antoniadis:1992xu}. This has recently been proved \cite{Komargodski:2011vj}. See  \cite{Shore:2016xor} for a review.

\item
Our discussions so far
are valid only for theories which are classically conformally invariant (e.g.~conformal scalars and massless fermions in any $D$, Maxwell/Yang-Mills in $D =4$, $p$-form gauge fields in $D=2p+2$, Conformal Supergravity in $D=(2,4,6)$). For other
theories  (e.g.~Maxwell/Yang-Mills for $D \neq 4$, pure quantum gravity for $D > 2$, or any
theory with mass
terms) the ``anomalies'' will still survive, but will be accompanied by
contributions to
$g^{\alpha\beta} \langle T_{\alpha\beta}\rangle$ expected anyway through the lack of
conformal invariance. Since 
the anomaly arises because the operations of regularizing and taking the trace
do not commute, the
anomaly ${\bf \cal A}$ in a theory which is not classically Weyl invariant may be defined as \cite{Duff:1977ay,Duff:1993wm,Casarin:2018odz}:
\begin{equation}
{\bf \cal A}~=g^{\alpha\beta} \langle T_{\alpha\beta} \rangle_{\tiny{\textrm{reg}}}-\langle g^{\alpha\beta}T_{\alpha\beta}\rangle_{\tiny{\textrm{reg}}}.
\label{A}
\end{equation}
Of course, the second term happens to vanish when the classical invariance is
present. 
That it still makes sense to talk of an anomaly in the absence of a symmetry is already familiar from the divergence of the axial 
vector current (\ref{kds}) where the operations of regularizing and taking the divergence do not commute
\be
\partial_{\mu}<(\sqrt{g}J^\mu{}_5)>_{reg}-<\partial_{\mu}(\sqrt{g}J^\mu{}_5)>_{reg}= -\frac{1}{384\pi^2}{}\epsilon^{\mu\nu\rho\sigma}R_{\alpha\beta\mu\nu}R^{\alpha\beta}{}_{\rho\sigma} 
\ee
and where the second term 
\be
<\partial_{\mu}(\sqrt{g}J^\mu{}_5)>_{reg}=2m<\sqrt{g}{\bar \psi}\gamma_5\psi>_{reg}
\ee
happens to vanish when the classical axial symmetry 
\be
\delta \psi=\theta\gamma_5\psi
\ee
is  present i.e. when the fermions are massless.
For theories which are not classically conformal, the Weyl anomaly and counterterms continue to be given  by the Schwinger-DeWitt  $B_4$ coefficient appearing in the asymptotic expansion of the heat-kernel   \cite{Hawking:1976ja,Christensen:1976vb,Dowker:1976zf} and equations (\ref{canda},\ref{n=4counterterm},\ref{pi}) get replaced by

\be
{\cal A}=\frac{1}{(4\pi)^2}(cF-aG+eR^2)
\ee
\begin{equation}
\Delta L=\frac{1}{\epsilon}\frac{1}{(4\pi)^2}\sqrt{g}(cF-aG+eR^2)
\label{n=4'counterterm}
\end{equation} 
\bea
\Pi_{\mu\nu\rho\sigma}(p)
&=&\frac{1}{\epsilon}\frac{1}{(4\pi)^2}[cS^{(2)}{}_{\mu\nu\rho\sigma}+ eS^{(0)}{}_{\mu\nu\rho\sigma}] +finite
\eea
where 
\be
S^{(0)}{}_{\mu\nu\rho\sigma}=\frac{1}{3}X_{\mu\nu} X_{\rho\sigma}
\ee
Note that the anomaly ${\cal A}$ is still local even though $g^{\alpha\beta} \langle T_{\alpha\beta} \rangle_{\tiny{\textrm{reg}}}$ and $\langle g^{\alpha\beta}T_{\alpha\beta}\rangle_{\tiny{\textrm{reg}}}$ separately need not be. ${\cal A}$ is not in general the functional derivative of any action, however, and is no longer constrained by the Wess-Zumino consistency condition. Hence the appearance of $R^2$.  The prescription (\ref{A}) is regularization-scheme independent but see \cite{Casarin:2018odz} for a recent and very clear articulation of this theme in the case of dimensional regularization.
\item
$p$-form gauge fields $\phi_p$ in $D\neq 2p+2$ provide nice examples of theories that are scale invariant but not conformal invariant. In $D=4$  $\phi_p$  and their duals $\phi_{(2-p)}$ yield  \cite{Duff:1980qv}
\be
\int {\bf \cal A}(\phi_2)-\int{\bf \cal A}(\phi_0)=\chi,
\ee
\be
\int {\bf \cal A}(\phi_3)=-2\chi,
\ee
 
Such {\it quantum inequivalence} of $p$-forms and their duals has been called into question \cite{Siegel:1980ax,Grisaru:1984vk,Bern:2015xsa} on the grounds that their {\it total} stress tensors are the same and that the anomalous trace is unphysical.  Nevertheless, the Euler number factors they provide in the partition functions are important for the subjects of Black Hole Entropy \cite{Bhattacharyya:2012ye}, Free Energy \cite{Raj:2016zjp} and Entanglement Anomalies \cite{Donnelly:2016mlc}.

\item
Another controversy concerns the role of the Pontryagin number \cite{Christensen:1978gi,Christensen:1978md,Townsend:1979js}. Consider the coefficients $B_k(A, B)$ with $(k = 0$ to $ \infty)$ which appear in the asymptotic expansion of the heat kernel corresponding to the operator $\Delta(A, B)$, the generalized Laplacian acting  on the $(A,B)$ representations of the (Eulideanized) Lorentz group  $SO(4)$  in $D=4$. It is this  coefficient $B_4$ which counts the sum of the number of zero-modes $n(A,B)$ and the number of non zero-modes $m(A,B)$  of $\Delta(A, B)$
\be
B_4(A,B) =n(A,B)+ m(A, B) 
\ee
As expected, if one calculates the axial anomaly via the difference of chiral and antichiral reps $(A,B)-(B,A)$ with $A\neq B$ the $\chi $ dependence drops out. Similarly, if we calculate the Weyl anomaly for the sum of chiral and antichiral reps $(A,B)+ (B,A)$, the $P$ dependence drops out. However (although this was not explicitly stated) the partition function for a purely chiral rep $(A,B)$ with $A\neq B$  depends on both $\chi$ and $P$ \footnote{The addition of both $\chi$ and $P$ to the action is discussed in \cite{Deser:1980kc}.}. This has been the subject of much debate recently with several papers arguing with different methods for the existence of a Pontryagin term in the Weyl anomaly \cite{Bonora:2018obr,Bonora:2019dyv,Nakagawa:2020gqc} while others disagree \cite{Bastianelli:2019zrq}.

\end{itemize}
\section{Subsequent developments}
\begin{itemize}
\item
Spacetime Weyl anomalies have found application in quantum corrections to the Schwarzschild solution and Newton's law  \cite{Duff:1974ud,Duff:2000mt}, particle creation \cite{Parker:1984rm}, the Hawking effect \cite{Christensen:1977jc}. inflationary cosmology \cite{Starobinsky:1980te}, asymptotic safety \cite{Weinberg:1980gg,Duff:1977ay}, wormholes \cite{Grinstein:1988eb}, holography \cite{Graham:1999pm,Henningson:1998gx}, viscosity bounds \cite{Baier:2007ix, Sinha:2009ev}, condensed matter physics \cite{Chernodub:2016lbo}, hydrodynamics \cite{Eling:2013bj}, the graviton mass in the braneworld \cite{Karch:2000ct}, conformal collider physics \cite{Hofman:2008ar}, quantum entanglement \cite{Nishioka:2009un, Casini:2011kv, Perlmutter:2015vma,Herzog:2015ioa, Donnelly:2016mlc}, log corrections to black hole entropy \cite{Sen:2011ba, Sen:2012cj,Astaneh:2014wxg}, generalized mirror symmetry \cite{Duff:2010ss, Duff:2010vy}, and double-copy theories \cite{Antoniadis:1992sa}.
\item
The cancellation of worldsheet Weyl anomalies not only determines  the critical dimensions $D=26$ for strings and $D=10$ for superstrings \cite{Polyakov:1981rd,Polyakov:1981re}, but also provides the derivation of the spacetime Einstein equations \cite{Callan:1985ia}.
\item
In \cite{Duff:1983vj} it was pointed out that Euclidean signature field configurations and their topological properties  (Betti numbers, Euler numbers, Pontryagin numbers, holonomy, index theorems etc) which feature in gauge and gravitational instanton physics
can lead a second life as {\it internal manifolds}  $X^n$ appearing in the compactification of the $n$ extra dimensions in Lorentzian signature Kaluza-Klein theory $M^D=M^4 \times X^n$. The first non-trivial example was provided by $K3$ \cite{Duff:1983vj}. 
\item
Moreover, the Weyl anomaly in supergravity, string and M-theory can depend also on the {\it internal} Euler number \cite{Duff:2010ss}.
In the case of Type IIA on $X^6$, for example,
\be
\int d^4x\sqrt{g}g^{\alpha\beta}\langle T_{\alpha\beta}\rangle =
-\frac{1}{24} \chi(M^4)\chi(X^6)=-\frac{1}{24} \chi(M^{10})
\label{n=10euler}
\ee
where
\be
 \chi(X^6)=2c_0-2c_1+2c_2-c_3
\ee
and $c_k$ are the betti numbers of $X^6$. Similarly for M-theory on $X^7$
\be
\int d^4x\sqrt{g}g^{\alpha\beta}\langle T_{\alpha\beta}\rangle =
-\frac{1}{24} \chi(M^4)\rho(X^7)
\label{n=11euler}
\ee
where
\be
 \rho(X^7)=7b_0-5b_1+3b_2-2b_3+b_4
\ee
and $b_k$ are the betti numbers of $X^7$. 
\end{itemize} 

 \section{\bf Memories of Transylvania}
 
In 2013 Peter Freund invited me to give the 7th annual Erwin Schr\"odinger Lecture at West University of Timisoara, Romania.  I also enjoyed the kind hospitality of Peter and his wife Lucy at their home, where he reminisced about his life as a student at the time of the 1956 anti-Soviet uprising. Identified as a rabble-rouser he faced the firing squad (a bit like Schr\"odinger's cat), but fortunately orders came to stand down before the triggers were pulled. Peter was a unique individual.
 I am glad to have inhabited a universe where both he and the cat survived. 
 
 \section{Acknowledgements}
 
 I am grateful to Leron Borsten and Silvia Nagy  for valuable discussions, to Philip Candelas for hospitality at the Mathematical Institute, University of Oxford, to Marlan Scully for his hospitality in the Institute for Quantum Science and Engineering, Texas A\&M University, and to the Hagler Institute for Advanced Study at Texas A\&M for a Faculty Fellowship. This work was supported in part by the STFC under rolling grant ST/P000762/1.

\bibliography{Weyl}
\bibliographystyle{utphys}
 
 \end{document}